\documentclass{iopart}

\newcommand{\kB}{{k_{\mathrm B}}}

\makeatletter
\long\def\@makefntext#1{\parindent 1em\noindent
 \makebox[1em][l]{\footnotesize\rm$\m@th{^\alph{mpfootnote}}$}%
 \footnotesize\rm #1}
\def\@makefnmark{%
\makebox[.5em][l]{\footnotesize\rm$\m@th{^\alph{mpfootnote}}$}%
}
\def\@thefnmark{\alph{footnote}}
\def\footnote{\@ifnextchar[{\@xfootnote}{\stepcounter{\@mpfn}%
       \begingroup\let\protect\noexpand
       \xdef\@thefnmark{\thempfn}\endgroup
     \@footnotemark\@footnotetext}}
\def\@xfootnote[#1]{\setcounter{footnote}{#1}%
   \addtocounter{footnote}{-1}\footnote}
\makeatother

\usepackage{graphicx}
 
\begin{document}
 
\title[Physical correlation in heavy Fermion superconductors]{$T_c$
   for heavy Fermion superconductors linked with other physical
   properties at zero and applied pressure}

\author{G. G. N. Angilella\dag, N. H. March\ddag\P, R. Pucci\dag}
\address{\dag\ Dipartimento di Fisica e Astronomia, Universit\`a di
   Catania, and\\
Istituto Nazionale per la Fisica della Materia, UdR Catania,\\
Via S. Sofia, 64, I-95123 Catania, Italy\\
\ddag\ Department of Physics, University of Antwerp,\\
Groenenborgerlaan 171, B-2020 Antwerp, Belgium\\
\P\ Oxford University, Oxford, UK}

\ead{Giuseppe.Angilella@ct.infn.it}

\begin{abstract}
The superconducting transition temperature $T_c$ has earlier been
   correlated with coherence length and effective mass for a series of
   heavy Fermion (HF) materials at atmospheric pressure.
Here, a further physical property, the dc~electrical conductivity
   $\sigma(T_c )$, is one focal point, another being the pressure
   dependence of both $T_c$ and $\sigma(T_c )$ for several HF
   materials.
The relaxation time $\tau(T_c )$ is also studied in relation to an
   Uncertainty Principle limit, involving only the thermal energy $\kB
   T_c$ and Planck's constant.
\end{abstract}

\submitto{\SUST}


\date{\today}

\section{Introduction}

In earlier work \cite{Angilella:00b,Angilella:03d}, the
   superconducting transition temperature $T_c$ of several heavy
   Fermion (HF) materials has been correlated with the effective mass
   $m^\ast$ (usually $\sim100m_e$, with $m_e$ the electron mass) and
   the coherence length $\xi$ by
\begin{equation}
\kB T_c = f \left( \epsilon_c \right),
\end{equation}
where $\epsilon_c$ is a characteristic energy defined as
   \cite{Angilella:00b}
\begin{equation}
\epsilon_c = \frac{\hbar^2}{m^\ast \xi^2} .
\label{eq:epsc}
\end{equation}
An approximate form of the relation between $\kB T_c$ and $\epsilon_c$
   has been derived with the 
   Bethe-Goldstone equation as starting point \cite{Angilella:03d}.
One finds
\begin{equation}
\frac{\epsilon_{\mathrm{F}}}{\epsilon_c} = \frac{4}{3} x^2 +
   \frac{\ell(\ell+1)}{1+x} \left( 1 - \frac{x\ln x}{1+x} \right),
\label{eq:anTc}
\end{equation}
where $2 \epsilon_{\mathrm{F}} x=|\epsilon| \simeq \kB T_c$ is the
   binding energy of a Cooper pair, and $\epsilon_{\mathrm{F}}$ is
   the Fermi energy \cite{Angilella:03d}.
Equation~(\ref{eq:anTc}) manifestly depends on the quantum number
   $\ell$ of the pair angular momentum, which is usually employed to
   parametrize the anisotropic character of the superconducting order
   parameter, $\ell=0,\, 1,\, 2$ corresponding to $s$-, $p$-, and
   $d$-wave symmetry, respectively.
While this expression correctly reduces to the standard one for
   isotropic $s$-wave superconductors, in the case $\ell>0$ it agrees
   qualitatively with the phenomenological dependence of $\kB T_c$ on
   the characteristic energy $\epsilon_c$, proposed in
   Ref.~\cite{Angilella:00b} for the HF compounds as well
   as for the high-$T_c$ cuprates.
It should be pointed out, however, that Equation~(\ref{eq:anTc}) is
   qualitatively insensitive to different values of $\ell >0$, the
   effect of a nonzero $\ell$ being mainly that of having $\kB T_c$
   saturating to a finite value as $\epsilon_c \to\infty$, instead of
   diverging, as is the case with $\ell=0$ \cite{Angilella:03d}.

Since this early work, we have uncovered in the literature further
   relevant data, \emph{e.g.} on CeCoIn$_5$, CeIrIn$_5$, and CeRh$_2$Si$_2$
   (cf. Table~\ref{tab:hf} and references therein).
Figure~\ref{fig:hf} then shows an updated correlation plot of $\kB
   T_c$ versus $\epsilon_c$, including the latter three new entries appearing
   at the two ends of the series of data, and with Eq.~(\ref{eq:anTc}) used
   as fitting function.

Motivated by the study of Homes \emph{et al.} \cite{Homes:04} on
   high-$T_c$ materials (plus elemental metals Nb and Pb with
   relatively high $T_c$ values for such superconductors) and of
   Zaanen \cite{Zaanen:04}, it is useful in connection with
   Table~\ref{tab:hf} to define a relaxation time $\tau (T_c )$
   through \cite{AM} 
\begin{equation}
\sigma (T_c ) = \frac{n_n e^2 \tau (T_c )}{m^\ast } \\
\label{eq:Zaanen}
\end{equation}
where $\sigma(T_c ) = \rho^{-1} (T_c )$ is the dc~electrical
   conductivity at the transition temperature $T_c$, and $n_n$ is the
   carrier density in the normal state. 
For several materials, experimental data collected in Table~\ref{tab:hf}
   for all the physical quantities
   appearing in Eq.~(\ref{eq:Zaanen}) exist with the exception of the
   relaxation time $\tau(T_c )$.
Table~\ref{tab:hf} therefore records the value of $\tau(T_c )$
   extracted from Eq.~(\ref{eq:Zaanen}) using experimental values for
   $\rho(T_c )$, $n_n$ and $m^\ast$.
For comparison, we have also recorded the `Uncertainty Principle' (UP)
   estimate $\tau_{\mathrm UP}$ given by Zaanen \cite{Zaanen:04},
   following the study of Homes \emph{et al.} \cite{Homes:04}:
\begin{equation}
\tau_{\mathrm UP} = \frac{\hbar}{\kB T_c} .
\label{eq:UP}
\end{equation}
In most cases, the values of $\tau(T_c )$ entered in
   Table~\ref{tab:hf} are of the same order of magnitude of
   $\tau_{\mathrm UP}$ given by Eq.~\eref{eq:UP}, but no simple
   correlation exists between $1/\tau(T_c )$ and $T_c$ in the HF
   materials.

\section{Pressure dependence of $T_c$ and $\rho(T_c )$}

\begin{figure}[t]
\centering
\includegraphics[height=0.9\columnwidth,angle=-90]{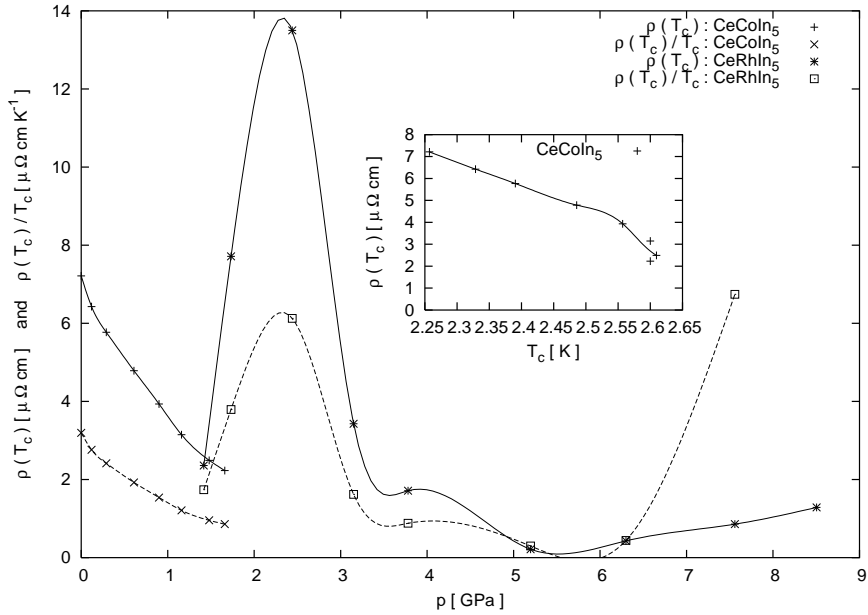}
\caption{Resistivity $\rho(T_c )$ at $T_c$ (solid lines) and ratio
   $\rho(T_c)/T_c$ (dashed lines) of CeCoIn$_5$ \cite{Nicklas:01} and
   CeRhIn$_5$ \cite{Muramatsu:01}. 
{\sl Inset:} Plots $\rho(T_c )$ versus $T_c$ for CeCoIn$_5$
   \cite{Nicklas:01}.
Lines are guides to the eye.}
\label{fig:pressure}
\end{figure}

Of the HF materials referred to above, we next note that CeCoIn$_5$,
   which is a superconductor at atmospheric pressure ($p=0$), has been
   studied over a pressure range out to about 3~GPa \cite{Nicklas:01}.
Figure~\ref{fig:pressure} plots the variation of the normal state
   resistivity $\rho(T_c )$, together with
   the ratio $\rho (T_c )/T_c$, as a function of pressure.
In contrast to the almost monotonic decrease of both these quantities
   with increasing pressure, we have also plotted available
   experimental data for CeRhIn$_5$ \cite{Muramatsu:01}, which however
   starts superconducting at 1.7~GPa.
The structure of both $\rho(T_c )$ and $\rho (T_c )/T_c$ as a function
   of pressure is marked, and they correlate.
The inset shows $\rho(T_c )$ versus $T_c$ constructed from the set of
   data for CeCoIn$_5$ \cite{Nicklas:01}.
Over a range of $T_c$ from 2.2 to 2.5~K, the behaviour is rather
   linear, followed by a more sudden decrease of resistivity with
   increasing $T_c$.

\section{Summary and future directions}

\begin{figure}[t]
\centering
\includegraphics[height=0.9\columnwidth,angle=-90]{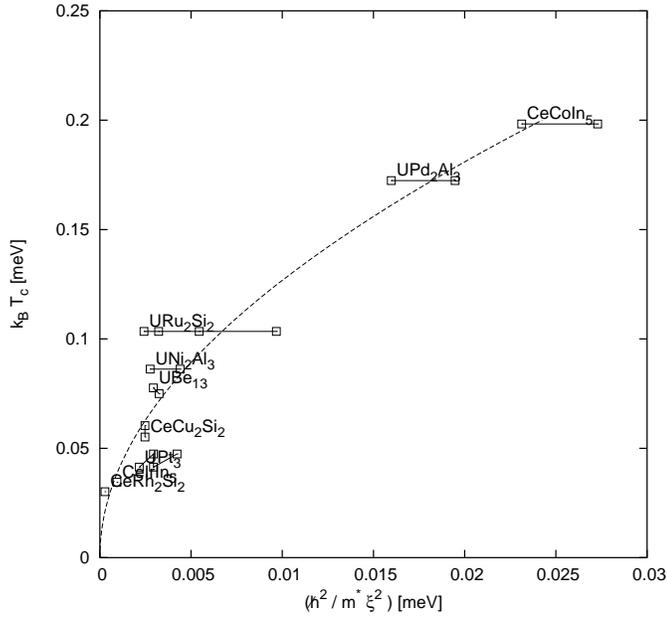}
\caption{Measured superconducting transition temperatures $T_c$ for a
   variety of HF materials plotted against characteristic
   energy $\epsilon_c$ defined in Eq.~(\ref{eq:epsc}).
Data for $T_c$, $m^\ast$, $\xi$ have been taken from
   Table~\ref{tab:hf}.
Dashed line is a phenomenological fit based on Eq.~(\ref{eq:anTc})
   \cite{Angilella:03d}.
}
\label{fig:hf}
\end{figure}

Our findings to this point may be summarized as follows.
The interlink between $T_c$ and the characteristic energy $\epsilon_c$ in
   Fig.~\ref{fig:hf} appears relatively robust, as evidenced by the
   addition of quite recent data.
The classification of HF materials is clearly quite different from the
   high-$T_c$ regularity plus Nb and Pb discussed by Homes \emph{et
   al.} \cite{Homes:04} and also by Zaanen \cite{Zaanen:04}.
However, the Uncertainty Principle relaxation time
   $\tau_{\mathrm{UP}}$ is found to be of the same order of magnitude
   as that extracted from measured dc~conductivity data plus effective
   masses at atmospheric pressure, namely $\tau (T_c )$.
However, no inverse correlation between $\tau (T_c )$ and $T_c$ is
   found for an admittedly limited number of HF materials.
In the same context, we have used the superconducting penetration
   depth $\lambda_0$ in Table~\ref{tab:hf} to estimate the superfluid
   density $\rho_s$ as $\lambda_0^{-2}$.
Then, following Homes \emph{et al.} \cite{Homes:04}, if we construct
   $\rho_s / T_c \sigma(T_c )$, then for all but one of the HF
   materials for which data is recorded in Table~\ref{tab:hf}, this
   ratio is at least a factor of 7 greater than for high-$T_c$
   materials, and with a huge scatter, confirming the above conclusion
   that HF materials are in a quite different category from high-$T_c$
   materials plus the elemental BCS superconductors Nb and Pb.

\begin{figure}[t]
\centering
\includegraphics[height=0.9\columnwidth,angle=-90]{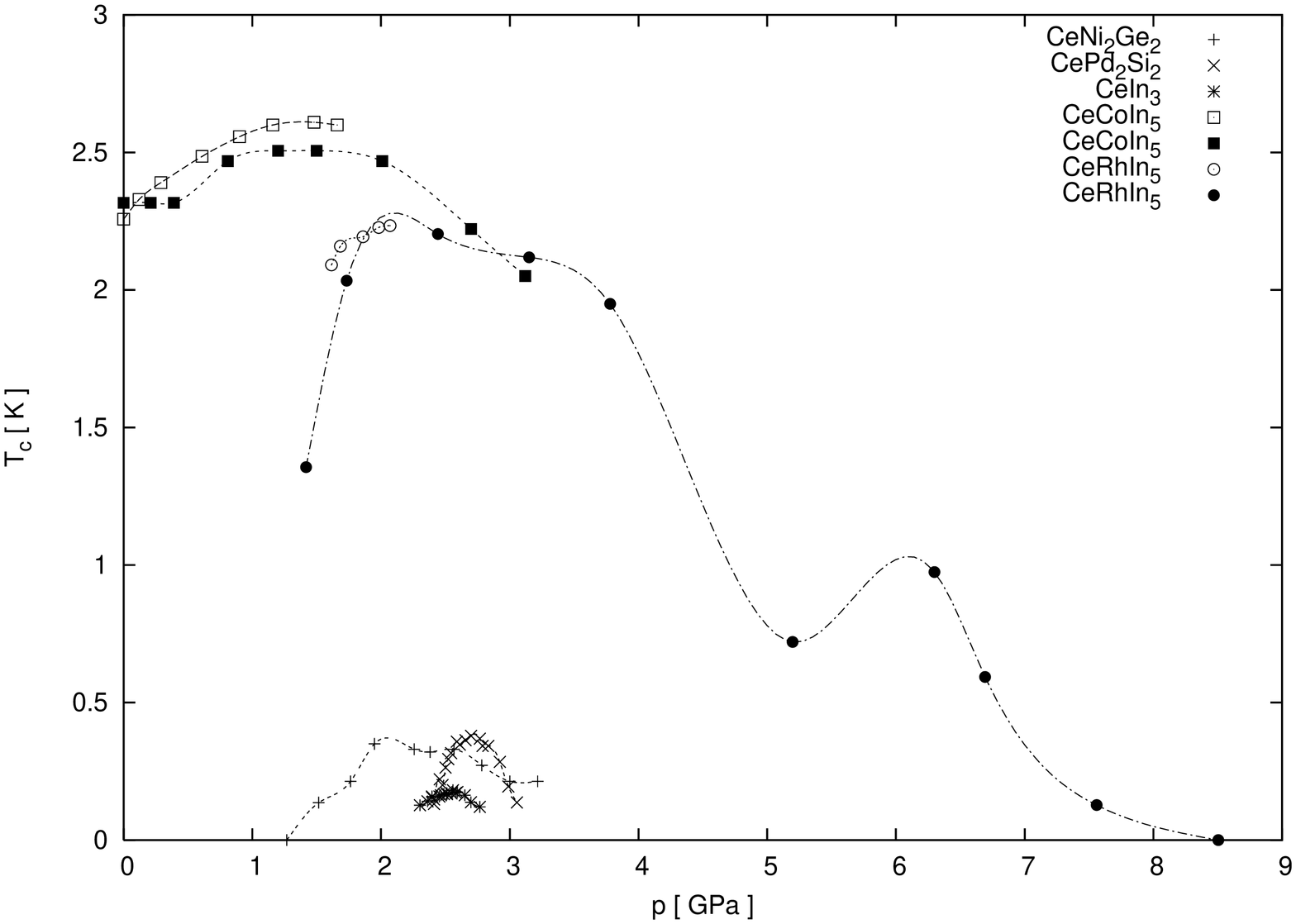}
\caption{Experimental superconducting transition temperatures $T_c$
   for CeNi$_2$Ge$_2$ ($+$~\cite{Grosche:00}), CePd$_2$Si$_2$
   ($\times$~\cite{Grosche:01}), CeIn$_3$ ($*$~\cite{Grosche:01}),
   CeCoIn$_5$ (\opensquare~\cite{Nicklas:01}, \fullsquare~
   \cite{Shishido:03}), CeRhIn$_5$ (\opencircle~\cite{Hegger:00},
   \fullcircle~\cite{Muramatsu:01}), plotted as a function of
   pressure $p$.
Lines are guides to the eye.
}
\label{fig:corr}
\end{figure}

As to future directions, we feel that further work, both on experiment
   and theory, by applying pressure to HF materials, should be
   illuminating.
Thus, we have collected in Figure~\ref{fig:corr} some available
   experimental data for $T_c$ as a function of pressure.
The simplest example, and the only one shown which superconducts at
   $p=0$, is CeCoIn$_5$ \cite{Nicklas:01,Shishido:03}, which has a
   relatively smooth variation of $T_c$ with pressure, exhibiting a
   single maximum.
It is tempting for the future to study whether a link can be forged,
   for $p<p_{\mathrm{max}}$, the latter corresponding to the maximum
   of $T_c$, with Fig.~\ref{fig:hf}.
However, whereas Fig.~\ref{fig:hf} relates $T_c$ to a single variable,
   \emph{i.e.} the characteristic energy $\epsilon_c$ defined in
   Eq.~(\ref{eq:epsc}), it may be that one must add further variables
   to describe the pressure dependence of $T_c$.
For example, the detail of spin fluctuation
   \cite{Monthoux:99,Abanov:01,Angilella:01a,Curro:03}, believed presently to
   be at least partially responsible for Cooper pair formation in this
   class of materials, may need inclusion.
However, of course, some account is already present through the
   coherence length $\xi$, in which the size of the Cooper pair is
   manifested.
Of course, for the remaining materials in Fig.~\ref{fig:corr}, the
   pressure dependence of $T_c$ is more complex, including the fact
   that pressure is needed already to induce superconductivity.

\begin{table}[t]
\caption{Selected physical properties for uranium and
   cerium based HF materials.
Where available, multiple entries separated by slashes refer to
   properties along different crystallographic directions.
$T_N$ is the magnetic ordering (N\'eel) temperature, $\gamma$ denotes
   the Sommerfeld specific-heat coefficient, $\lambda_0$ the
   superconducting penetration depth extrapolated at $T=0$, and
   $\omega_{pn}$ is the plasma frequency in the normal state.
The last two columns are the `Uncertainty Principle' relaxation time
   $\tau_{{\mathrm UP}}$, Eq.~\eref{eq:UP}, and the relaxation time $\tau
   (T_c )$ at $T_c$, Eq.~\eref{eq:Zaanen} \cite{Zaanen:04}. 
}
\label{tab:hf} 
\begin{minipage}{\textwidth}
\begin{tabular}{lccccc}
\br
Compound & $T_c$ & $T_N$ & $\xi$ &
$m^\ast / m_e$ & $\gamma$ \\
 & (K) & (K) & (\AA) & & (J~mol$^{-1}$~K$^{-2}$) \\
\br
UPt$_3$ & 0.52, 0.48 
   \cite{Amato:97} & 5.0 \cite{Heffner:96, Amato:97} 
   & 100/120 
   \cite{Heffner:96} & 180 \cite{Heffner:96} & 0.450 \cite{Heffner:96}
   \\
UBe$_{13}$ & 0.87 \cite{Amato:97} & --- & 100 \cite{Heffner:96} & 260
   \cite{Heffner:96} & 1.100 \cite{Heffner:96} \\
UNi$_2$Al$_3$ & 1.0 \cite{Heffner:96,Amato:97} & 4.3$-$4.6
   \cite{Heffner:96,Amato:97} & 240 
   \cite{Heffner:96} & 48 \cite{Heffner:96} & 0.120 \cite{Heffner:96}\\ 
UPd$_2$Al$_3$ & 2.0 \cite{Heffner:96,Amato:97} & 14.5 \cite{Amato:97} & 85
   \cite{Heffner:96} & 66 \cite{Heffner:96} & 0.145 \cite{Heffner:96}\\
URu$_2$Si$_2$ & 1$-$1.5 \cite{Amato:97,Degiorgi:99} & 17$-$17.5
   \cite{Heffner:96,Degiorgi:99} & 100/150 
   \cite{Heffner:96} & 140 \cite{Heffner:96} & 0.065$-$0.18
   \cite{Heffner:96,Degiorgi:99}\\
\bs
CeCu$_2$Si$_2$ & 0.65 \cite{Amato:97} & 1.3 \cite{Heffner:96} & 90
   \cite{Heffner:96} & 380 \cite{Heffner:96} & 0.73$-$1.1
   \cite{Heffner:96}\\ 
CeRh$_2$Si$_2$ & 0.35 \footnote{At 0.9~GPa.} \cite{Movshovich:96} & 35$-$36
   \cite{Movshovich:96,Alsmadi:03} & 370 \cite{Movshovich:96} & 220
   \cite{Movshovich:96} & 0.08 \cite{Movshovich:96}\\ 
CePd$_2$Si$_2$ & 0.4 \footnote{At 2.71~GPa.} \cite{Grosche:01} & 10.2
   \footnote{At 0~GPa. $T_N$ vanishes at $p_c = 2.86$~GPa \cite{Grosche:01}.}
   \cite{Grosche:01} & 150 \cite{Grosche:01} & & 0.13 \cite{Lister:97}\\ 
CeCu$_2$Ge$_2$ & 0.64 \footnote{At 10.1~GPa.} \cite{Jaccard:92} & 4.1
   \cite{Jaccard:92} & & & \\
CeNi$_2$Ge$_2$ & 0.22 \footnote{At 1.5~GPa.} \cite{Lister:97} & & & &
   0.4 \cite{Grosche:01} \\ 
CeRu$_2$Ge$_2$ & 7.40 \cite{Wilhelm:04} & 8.55 \cite{Wilhelm:04} & & &
   \\
\bs
CePt$_3$Si & 0.75 \cite{Bauer:04} & 2.2 \cite{Bauer:04} & 81$-$97
   \cite{Bauer:04} & & 0.39
   \cite{Bauer:04}\\ 
CeNiGe$_2$ & --- & 3, 4 \cite{Jung:02,Alsmadi:03} & --- & & 0.22
   \cite{Jung:02}\\ 
CeNiGe$_3$ & 0.48 \footnote{At 4$-$10~GPa.} \cite{Nakashima:04} & 5.5
   \footnote{At 0~GPa. $T_N$ vanishes at $p_c = 5.5$~GPa \cite{Nakashima:04}.}
   \cite{Nakashima:04} & 130 \footnote{At 6.5~GPa.}
   \cite{Nakashima:04} & & 0.034 \cite{Nakashima:04}\\ 
\bs
CeCoIn$_5$ & 2.3 \cite{Petrovic:01a} & & 58 \cite{Movshovich:01}, 35/82
   \cite{Settai:01} & 83 \cite{Movshovich:01}, 5/49/87
   \cite{Settai:01} & 0.29 \cite{Petrovic:01a,Movshovich:01}\\
CeRhIn$_5$ & 2.1 \footnote{At 1.7~GPa. $T_c$ reaches 2.2~K at 2.5~GPa
   \cite{Muramatsu:01}.} \cite{Hegger:00} & 3.8 \cite{Hegger:00} & 57
   \footnote{At 2.5~GPa.} \cite{Muramatsu:01} & & 0.40 \cite{Hegger:00}\\ 
CeIrIn$_5$ & 0.40 \cite{Petrovic:01} & 0 \cite{Zheng:04} & 241
   \cite{Movshovich:01} & 140 \cite{Movshovich:01}, 20/30 \cite{Haga:01}
   & 0.72$-$0.75 \cite{Petrovic:01,Petrovic:01a}\\
\bs
CeIn$_3$ & 0.25 \footnote{At 2.5~GPa.} \cite{Hegger:00} &
   10 \footnote{At 0~GPa.}
   \cite{Hegger:00} & & & 
   $\leq 0.13$ \footnote{At 0~GPa.}
   \cite{Nasu:71,Nicklas:01}\\  
CePd$_3$ & --- & & --- & 36 \cite{Degiorgi:99} & 0.037 \cite{Degiorgi:99}\\
\br
\end{tabular}
\end{minipage}
\end{table}

\setcounter{table}{0}

\begin{table}[h]
\caption{(Continued)}
\begin{minipage}{\textwidth}
\begin{tabular}{lcccccc}
\br
Compound & $\rho(T_c )$ & $\lambda_0$ & $\omega_{pn}$ &
$n_n$ & $\tau_{{\mathrm UP}}$ & $\tau (T_c )$ \\
 & ($\mu\Omega$~cm) & (\AA) & (eV) & ($10^{22}$~cm$^{-3}$) & (ps) & (ps) \\
\br
UPt$_3$ & 0.3$-$3 \cite{Heffner:96} & $>15000$ \cite{Amato:97} & 2.6
   \cite{Degiorgi:99} & 1.8 
   \cite{Sulewski:88} & 13.9 & 11.8$-$118 \\
UBe$_{13}$ & 18 \cite{Heffner:96} & 11000 \cite{Amato:97} & & & 8.8 & \\
UNi$_2$Al$_3$ & 7 \cite{Heffner:96} & 3300 \cite{Amato:97} & 2.9
   \cite{Cao:96} & 0.61 \cite{Cao:96} & 7.6 & 4.0 \\ 
UPd$_2$Al$_3$ & 4 \cite{Heffner:96} & 4000 \cite{Amato:97} & 5.5
   \cite{Dressel:00} & 1.9 \cite{Degiorgi:94} & 3.8 & 3.1 \\
URu$_2$Si$_2$ & 12$-$70 \cite{Heffner:96} & 10000 \cite{Amato:97} & & &
   5.1$-$7.6 & \\ 
\bs
CeCu$_2$Si$_2$ & 2$-$65 \cite{Heffner:96} & 5000 \cite{Amato:97} & & 11.4
   \cite{Movshovich:96} & 11.8 & 0.2$-$6.0 \\ 
CeRh$_2$Si$_2$ & 2 \cite{Movshovich:96} & & & 19.7 \cite{Movshovich:96}
   & 21.8 & 2.0 \\  
CePd$_2$Si$_2$ & 1.4 \cite{Grosche:01} & & & & 19.1 & \\ 
CeCu$_2$Ge$_2$ & $\approx 6$ \cite{Jaccard:92} & & & & 11.9 & \\
CeNi$_2$Ge$_2$ & $\approx 3$ \cite{Grosche:01} & & & & 34.7 & \\ 
CeRu$_2$Ge$_2$ & & & & & 1.0 & \\
\bs
CePt$_3$Si & 6.5 \cite{Bauer:04} & & & & 10.2 & \\ 
CeNiGe$_2$ & --- & --- & & & --- & --- \\ 
CeNiGe$_3$ & & & & & 15.9 & \\ 
\bs
CeCoIn$_5$ & 7.21 \cite{Nicklas:01} & 2810 \cite{Ozcan:03} & & 1.15
   \cite{Movshovich:01} & 3.3 & 3.55 \\
CeRhIn$_5$ & 5$-$7.5 \cite{Hegger:00,Muramatsu:01} & & & & 3.6 & \\ 
CeIrIn$_5$ & $\leq 1$ \cite{Petrovic:01,Movshovich:01} & & & 2.67
   \cite{Movshovich:01} & 19.1 & 18.6 \\
\bs
CeIn$_3$ & & & & & 30.6 & \\  
CePd$_3$ & --- & --- & 2.3 \cite{Degiorgi:99} & & --- & --- \\
\br
\end{tabular}
\end{minipage}
\end{table}

\clearpage

\ack

We thank N. J. Curro, G. Sparn, and H. Wilhelm for helpful discussions
   and correspondence. 
NHM wishes to acknowledge that his contribution to this study was
   brought to fruition during a visit to the University of Catania in
   2004.
He wishes to thank the Department of Physics and Astronomy for the
   stimulating atmosphere and for generous hospitality.

\section*{References}
 
\bibliographystyle{sust}
\bibliography{a,b,c,d,e,f,g,h,i,j,k,l,m,n,o,p,q,r,s,t,u,v,w,x,y,z,zzproceedings,Angilella}

\end{document}